\documentclass[aps,prb,twocolumn,groupaddress,showpacs,showkeys]{revtex4-1}
\usepackage{amsmath,amssymb}
\usepackage{graphicx,epsfig}
\usepackage{txfonts}
\usepackage{color}
\usepackage{hyperref}
\usepackage{natbib}
\usepackage[utf8]{inputenc}
\usepackage{bbold}
\usepackage{relsize}

\begin{document}

\title{Influence of Multiplicative Stochastic Variation on Translational Elongation Rates}
\author{Sandip Datta and Brian Seed}
\email[]{bseed@ccib.mgh.harvard.edu}
\affiliation{Center for Computational and Integrative Biology, Massachusetts General Hospital, Boston, USA}
\date{\today}

\begin{abstract}
Recent experiments have shown that stochastic effects exerted at the level of translation contribute a substantial portion of the variation in abundance of proteins expressed at moderate to high levels. This study analyzes translational noise arising from fluctuations in residue-specific elongation rates. The resulting variation has multiplicative components that lead individual protein abundances in a population to exhibit approximately log-normal behavior. The high variability inherent in the process leads to parameter variation that has the features of a type of noise in biological systems that has been characterized as “extrinsic.” Elongation rate variation offers an accounting for a major component of extrinsic noise, and the analysis provided here highlights a probability distribution that is a natural extension of the Poisson and has broad applicability to many types of multiplicative noise processes.
\end{abstract}

\maketitle
\section{Introduction}
\label{sec:Introduction}

Regulation of the abundance of proteins expressed in living cells is mediated by multiple types of control, exerted over the rates of transcription, post-transcriptional mRNA processing, mRNA decay, translation, and protein degradation. The regulatory processes can be construed as a sequence of chemical reactions in a domain in which the number of participating molecules is small and hence stochastic influences are significant. Those influences give rise to fluctuations in protein concentration in an otherwise homogeneous cell population at steady state \citep{Bar-Even2006}.

Stochastic fluctuations in protein distribution can result in heterogeneous phenotypes in clonal populations that can be beneficial for the survival of a population of organisms in a changing environment \citep{BalAextexclamdownzsi2011}. For example, under conditions of nitrogen limitation, cyanobacteria dedicate a subpopulation of cells to nitrogen fixation while the rest of the population remains phototrophic \cite{Wolk1996}. Similarly, following exhaustion of nutrient resources, the undifferentiated free-living amoebae of cellular slime molds aggregate and undergo spontaneous differentiation into spore- and stalk-forming cells.  Such task-sharing decisions assisted by stochastic differentiation in a clonal population may have formed the basis for multi-cellular development \cite{BalAextexclamdownzsi2011, Beaumont2009}. In mammalian cells, the manifestation of stochastic gene expression resulting in phenotypic diversity has been observed in the processes of cellular differentiation \citep{BalAextexclamdownzsi2011} and apoptosis \cite{Spencer2009}.  Genes belonging to the same functional group have been found to possess similar noise characteristics \citep{Bar-Even2006, Newman2006}, and stress response genes that mitigate the effects of environmental fluctuations have been found to exhibit noisier expression than genes thought to require invariant expression \citep{Bar-Even2006, Newman2006}.

\section{Intrinsic and extrinsic noise and stochastic variation in protein abundance}
\label{sec:Intrinsic_extrinsic}

Elowitz et. al. have differentiated between intrinsic and extrinsic noise \citep{Elowitz2002}.  Sources of intrinsic noise include the random birth and death of molecules, or stochastic gene activation \citep{Bar-Even2006, Paulsson2004, Maheshri2007}. Extrinsic sources of noise include factors contributing to fluctuations in reaction rates \citep{Paulsson2004}, e.g.  the number of RNA polymerase molecules and ribosomes \citep{Raj2008a}, the variation in kinetic parameters such as rates of transcription and translation \citep{Bar-Even2006, Maheshri2007}, the variations in individual cell shapes and volume \citep{Maheshri2007, Raj2008a, Volfson2006}, and variation in common elements upstream of a transcription factor \cite{Volfson2006}.   

For intrinsic sources the resulting noise (normalized by the squared mean abundance) is inversely proportional to the mean protein abundance \citep{Paulsson2004},  and deviations from this relation help to identify the relative proportion of extrinsic noise. Detailed experimental measurements carried out in \textit{S. cerevisiae} have shown that the contribution of extrinsic noise to protein abundance increases with level of expression \citep{Bar-Even2006, Newman2006}. These findings are consistent with conclusions from earlier studies   \citep{Elowitz2002, Blake2003, Raser2005} that the source of noise for moderately to highly abundant proteins in both \textit{E. coli} and \textit{S. cerevisiae} is primarily extrinsic \citep{Maheshri2007}.  

Xie and coworkers have carried out single molecule measurements in \textit{E. coli} cells under conditions in which expression is highly repressed, so that the random formation and degradation of RNA molecules is the dominant noise source \citep{Cai2006, Yu2006}. They found that, below ten proteins per cell the noise is inversely proportional to protein abundance \citep{Taniguchi2010}. They also showed that the distribution of low copy number proteins can be fit to a gamma distribution, the two parameters of which have direct physical interpretation as the protein burst rate and burst size \citep{Taniguchi2010, Li2011}. Above ten proteins per cell the noise reaches a plateau indicating the dominance of extrinsic noise \citep{Taniguchi2010, Li2011}.  Reported numbers of proteins per bacterial cell can range from approximately 50,000 to nearly zero \citep{Huang2007, Malmstrom2009}. 

Two general models for the influence of transcriptional noise have been proposed, the Poisson and the telegraph processes. Under a Poisson process transcription occurs with constant probability in time resulting in single mRNAs being produced and destroyed \citep{Paulsson2004}. In a telegraph process the genes switch between transcriptionally active or inactive states and the active state results in a burst of mRNA production \citep{Chubb2006, Kaufmann2007}. In both processes the mRNA noise variance is inversely proportional to the mean mRNA abundance, but the proportionality constant for the Poisson process is one, and greater than one for the telegraph process \citep{Golding2005, Raj2009}. Higher eukaryotes exhibit a much broader mRNA distribution than prokaryotes and show transcriptional bursts \citep{Raj2006}.  A growing body of work suggests that transcription may occur in any single organism within a range of kinetic modes, a subset of genes being transcribed by Poisson processes, and a subset being transcribed with differing bursting dynamics \citep{Chubb2006, Larson2009, Suter2011}.

The correlation between mRNA and protein abundances (or lack thereof) can be used as a guide to the extent by which fluctuations in mRNA copy number produce fluctuations in protein concentration \citep{Raj2009}.  Previous studies have found that the correlation between mRNA and protein levels is poor across all organisms \citep{SousaAbreu2009, Maier2009, Schwanhausser2011}.  These studies have broadly indicated that post-transcriptional effects determine steady state protein abundance \citep{Vogel2012}.  Recently, Schwanaeusser et al. measured mRNA and protein simultaneously for 5000 genes in mouse fibroblasts and found that about $55\%$ of the correlation between mRNA and protein level can be explained by considering translation rate constants alone \cite{Schwanhausser2011}. Therefore as anticipated earlier \citep{Bar-Even2006, Maheshri2007, Larson2009}, among post-transcriptional steps, translation represents the most consequential stochastic factor for determining protein abundance for the cell as a whole. Cells expend more energy in translation compared to transcription (in an approximately 9:1 ratio), which may explain the dominance of translational control. 

\section{Translation proceeds at a variable rate}
\label{sec:Translation_variableRate}

Translation can be divided into four stages: initiation, elongation, termination and recycling. The mechanisms for initiation and termination differ between prokaryotes and eukaryotes, but the elongation mechanism is conserved \citep{Kapp2004}. Elongation is frequently the rate-limiting process for protein synthesis \citep{Wohlgemuth2011}. In the elongation phase, a series of reaction steps leads to the accommodation and addition of amino acid residues to the polypeptide chain (or rejection of the aa-tRNA), and each of these reactions can be characterized in terms of kinetic rate constants \citep{Rodnina2001}. As a simplification, a net effective rate constant for a single residue addition in elongation can be composed from the rate constants for individual steps that result in chain extension. The effective kinetic rate constant is expected to fluctuate from cell to cell across a cell population, depending on a variety of  noise sources, the vast majority of which will arise from proteins that compose the translational machinery and are expressed at a high level and hence are expected to contribute extrinsic noise.

In bacteria the elongation rate varies between 4 and 22 amino acids per second \citep{Wohlgemuth2011}. The protein sythesis rate can be affected by many factors, of which the most significant is considered to be the relative concentration of various tRNAs \citep{Ikemura1981, Varenne1984}. At each elongation step the ribosome must intercept the aminoacyl-tRNA (aa-tRNA) complimentary to the codon at the ribosome A site. The relative local concentration of various aa-tRNAs near the site determines the waiting time.  Codons corresponding to under-represented tRNAs reduce the elongation rate \citep{Varenne1984, SAcrensen1989} and are themselves under-represented \citep{Sharp1987}, which is thought to provide a mechanism allowing organisms to manipulate the expression level of proteins \citep{Makrides1996}. Elongation is also slowed by mRNA secondary structures called pseudoknots \citep{Namy2006} or by the interaction of nascent  peptide sequences with the ribosome exit channel \citep{Nakatogawa2002}.  Recently,  Ignolia et al.  have extensively sequenced ribosome protected mRNA fragments  thereby obtaining a more detailed picture of the ribosome distribution on mRNA \citep{Ingolia2011}. They observed substantial variation in the density of ribosome  footprints along mRNAs in both yeast and E. coli \citep{Ingolia2009}. In mammalian cells, some locations on mRNA were found to have 25-fold greater density than the median density across the gene \citep{Ingolia2011}.  Thousands of such sites were observed in mouse embryonic stem cell transcripts \citep{Ingolia2009}. Similar translational pauses have been reported \cite{Wolin1988, Darnell2011}. Current experimental procedures, including ribosome profiling, cannot provide the duration of such stochastic translational pauses , which are typically transient \cite{Shoemaker2012}. Additional factors affecting elongation rate are collisions between individual ribosomes in polysomes \citep{Mitarai2008}, controlled ribosome stalling, and interactions between the translating ribosome and RNA polymerase in prokaryotes \citep{Proshkin2010}.

\begin{figure}[t]
\begin{centering}
\includegraphics[width=1.0\columnwidth]{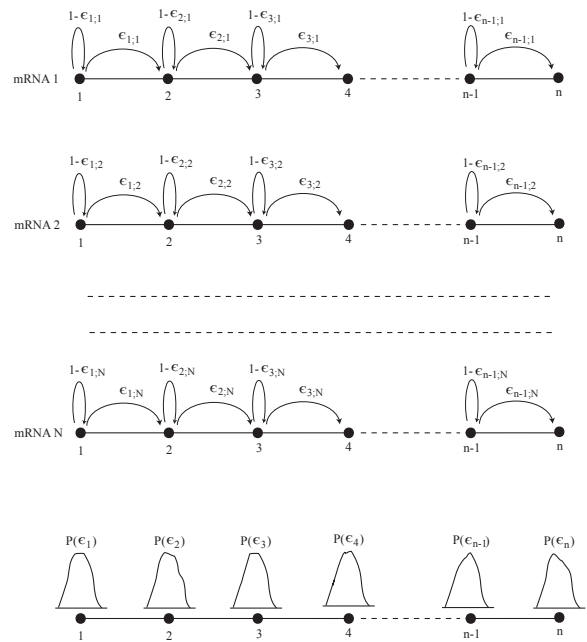} 
\par
\end{centering}
\setlength{\abovecaptionskip}{-30pt plus 3pt minus 2pt}
\caption{\label{fig:ModelSketch} Schematic diagram of the stochastic random walk model for the translation of mRNAs. Following initiation the translation proceeds via elongation at different local rates on different mRNAs.  The figure shows N copies of mRNA in a population of $N_c$ cells ($N>N_c$) undergoing elongation. Each mRNA is represented by a linear discrete lattice chain with individual nodes representing codons at which ribosomes add residues to polypeptides. At any given node i in a microscopic interval of time, the probability of a codon being translated or not is given by $\epsilon_i$ and $1-\epsilon_i$ respectively. The distribution of $\epsilon_i$ at the $i$th node is given by $P(\epsilon_i)$. $\epsilon_{i;j}$ represents the scaled rate constant (drawn from probability distribution $P(\epsilon_i))$ for jth mRNA at the $i^{th}$ site. Once the ribosome reaches site $n$, the termination site, a completed protein is released and the ribosome moves back to the initiation site (the recycling step).}
\end{figure}

Protein synthesis kinetics also depend on macroscopic factors, such as the overall metabolic status of the cell, composition of the template pool, the fraction of synthesis devoted to secreted versus non-secreted proteins, and the density of ribosomes on the template. The quantities of EF-G \citep{Dorner2006}  and elF5A \citep{Saini2009} significantly influence the rate at which elongation proceeds. Recent modeling of the dynamics of protein synthesis has focused on analysis of conditions for which inter-ribosome interaction is significant and has emphasized situations in which the movement of the ribosome is controlled by the availability of adjacent free mRNA exposed by the departure of the preceding ribosome \citep{Zouridis2007}. These studies emphasize the importance of the rate of motion of the leading ribosome in a transcript in determining the  number of copies per transcript. The formation of the mature protein by cotranslational protein folding is regulated by the modulation in the rate at which amino acids are added to the chains, which results from the stochastic variations in elongation rates \citep{OBrien2012}.

Cultures pulse-labeled with radioactive amino acids have been reported to exhibit a pattern of discrete intermediates corresponding to incomplete polypeptide chains, in which the intermediates can be visualized as bands on a polyacrylamide gel \citep{Varenne1984}. In the case of the abundantly expressed protein colicinA, a reasonably good fit could be made to the predicted rate of elongation based on the abundance of charged amino-acyl tRNAs for the known codons \citep{Varenne1984}. 

\section{A stochastic model for the translation process}
\label{sec:Translation_model}

A fraction of the elongation rate noise is codon-specific, resulting in different codons being translated at different rates \citep{Rodnina2001, Gromadski2004, Fluitt2007}. Gromadski and Rodnina measured the rate constants for different kinetic substeps for the CUC codon \citep{Gromadski2004}. Based on their data, Fluitt et al. made an estimate of the possible translation rates for all codons \citep{Fluitt2007}. For any particular  translating mRNA in a collection of cells, the rate constant at any arbitrary location along the mRNA is a stochastic variable and therefore can be described most appropriately by a distribution. A general analysis of elongation rates should take into account the propensity for eukaryotic ribosomes to undergo reinitiation following completion of translation, a phenomenon that is physically visualized under conditions of high protein synthesis as a circular template structure \citep{Philipps1965, Christensen1987}. With this consideration, the stochastic movement of a ribosome along the mRNA during elongation under the collective influence of various noise sources can be modeled as a unidirectional random walk on a chain with circular boundary conditions (see Figure~\ref{fig:ModelSketch}), with each incorporation of a residue taken to follow first order kinetics, with rate constant $\epsilon_i$ for the $i$-th residue. A reinitiation probability, $\lambda$, conveys the likelihood of reinitiation once a ribosome has reached the end of the open reading frame of $d$ codons.

The evolution of the ribosome motion for given initial conditions is determined in the usual manner by the exponentiation of a transition matrix. The lapse of an interval of time,$t$,  results in a change in the probability state vector $V_i (t)=Q_{ij} (t) V_j (0)$ for the leading ribosome position determined by the initial conditions and
\begin{equation}
\label{eqm1}
Q(t)=e^{-\alpha t} \sum_{k=0}^{\infty} \frac{(\alpha t)^k}{k!} U^{k}=e^{\alpha t(U-1)}
\end{equation}
where $U$ is a stationary transition matrix of the length of the polypeptide, $d$, having generator $T=U-\mathbb{1}_{d}$ where $\mathbb{1}_{d}$ is the unit matrix of length $d$. For the circular boundary conditions characteristic of eukaryotic elongation, the $i,j$ entry of the exponentiation of this matrix for  $i < j$ yields
\begin{equation}
\label{eqm2}
Q(t)_{i,j}=\frac{1}{2\pi i}\oint\frac{e^{ts}\lambda\prod_{k=i+1}^{j-1}(s+\epsilon_k)  \prod_{m=j}^{d} \epsilon_m \prod_{m=1}^{i-1}\epsilon_m}{\prod_{k=1}^{d}(s+\epsilon_k)-\lambda \prod_{m+1}^d \epsilon_m}ds
\end{equation}
and for $i\geq j$ yields
\begin{equation}
\label{eqm3}
Q(t)_{i,j}=\frac{1}{2\pi i}\oint\frac{e^{ts}\prod_{k=1}^{j-1}(s+\epsilon_k)  \prod_{k=i+1}^{d} (s+\epsilon_k) \prod_{m=j}^{i-1}\epsilon_m}{\prod_{k=1}^{d}(s+\epsilon_k)-\lambda \prod_{m+1}^d \epsilon_m}ds
\end{equation}
evaluated so that the contour encircles all the poles of the integrand, or, equivalently, encircles the pole at infinity in the opposite sense (see Appendix A). For large $d$ the product $\prod_{m=1}^{d} \epsilon_m$  is close to zero, and hence the roots of the denominator polynomial are expected to lie in the vicinity of $s=-\epsilon_k$. 

This picture is simplified in the case of short lived mRNAs or in the prokaryotic context, in which the contribution of ribosome recycling can be ignored. Setting $\lambda=0$ in Eq.~(\ref{eqm2}) and Eq.~(\ref{eqm3}) the elements of exponentiated matrix above reduce to 

\begin{align}
\label{eqm4}
&Q(t)_{i,j} = \nonumber \\
&\begin{cases} 0 & i<j \\ (-1)^{i+j}\prod_{k=j}^{i-1}\epsilon_k \sum\limits_{m=j}^{i} \frac{\mathlarger{e^{-t\epsilon_m}}}{\mathlarger{(\prod_{p=1}^{i-m}\epsilon_m - \epsilon_{m+p})(\prod_{q=j}^{m-1} \epsilon_m - \epsilon_q)}} & i\geq j \end{cases}
\end{align}

The essential element of formulas Eq.~(\ref{eqm2}) to Eq.~(\ref{eqm4}) from the standpoint of stochastic structure is the presence of high order products of the random variables $\epsilon_m$. When the $\epsilon_m$ are equal Eq.~(\ref{eqm4}) simplifies to the familiar Poisson:

\begin{equation}
\label{eqm5}
Q(t)_{i,j} = \begin{cases} 0 & i<j \\ \mathlarger{e^{-t \epsilon}\frac{(t\epsilon)^{i-j}}{(i-j)!}} & i\geq j \end{cases}
\end{equation}

Thus the discrete distribution $Q(t)_{i,j}$ represents a generalization of the Poisson that incorporates multiplicative stochastic variation and is appropriate for the characterization of processes that involve discrete steps that are subject to inter-step variability. Both transcription and translation are such processes, although the focus of this work is translation. Translational variation is likely to be greater than transcriptional variation because of the greater variety of participating substrates and the larger number of discrete steps that must occur to effect the addition of a single residue to the elongating chain. 

\begin{figure}[t]
\begin{centering}
\includegraphics[width=1.0\columnwidth]{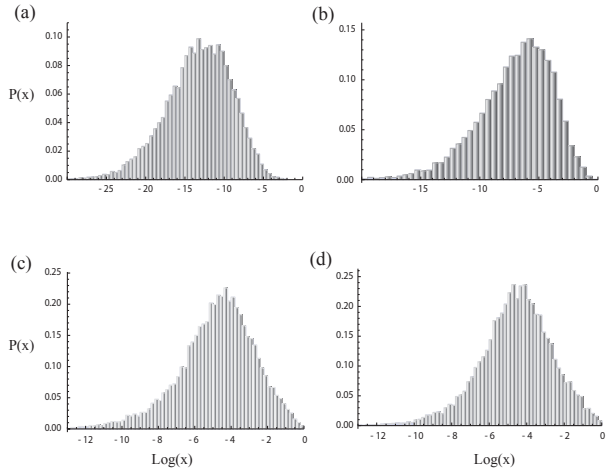} 
\par
\end{centering}
\setlength{\abovecaptionskip}{-50pt plus 3pt minus 2pt}
\caption{\label{fig:SteadyState} Simulation results for the distribution of proteins across cell population as the system evolves from an initial transient state and finally reaches the steady state. The plots give the distributions at times (a) $t = 0.5T$, (b) $t = T$, (c) $t = 2T$ and (d) $t = 3T$, where $T$ is an arbitrary unit of time. The distribution narrows between time $0.5T$ and time $T$ while the dynamics are in the transient phase. The narrowing continues as the dynamics near steady state at time $2T$. At time $3T$ a steady state distribution given by a log-normal is reached, confirmed by the finding that for any further increase in time the distribution remains invariant and log-normal. In the simulation ribosome recycling is incorporated via circular boundary condition. The distribution of completed proteins is calculated by taking the difference of fluxes between the termination and initiation sites.}
\end{figure}

\begin{figure}[t]
\begin{centering}
\includegraphics[width=1.0\columnwidth]{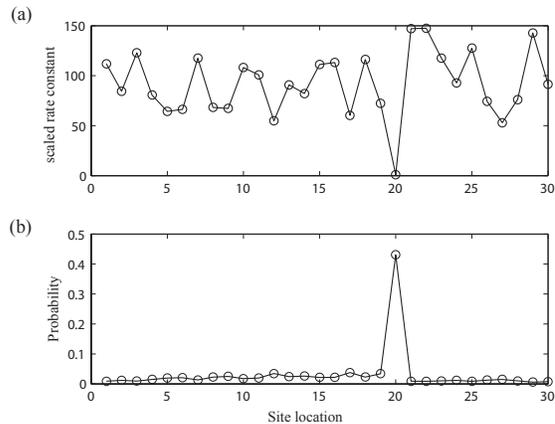} 
\par
\end{centering}
\setlength{\abovecaptionskip}{-50pt plus 3pt minus 2pt}
\caption{\label{fig:RibosomePausing} The effect of a low rate constant for elongation at a given residue. In this case it is expected that the elongation will stall at the site with low rate constant. (a) In order to simulate this condition the mean value for the rate constant distributions at various sites is allowed to vary between 50 and 150 except at site 20, where the mean value is set near zero. (b) The resulting distribution of the polypeptides across various sites clearly shows the stalling effect with resulting accumulation around the residue site 20. Although this example is artificially constructed to verify the validity of the model described in the present work, the occurrence of pause sites has been reported in the literature.}
\end{figure}

\section{Steady state distribution of proteins}
\label{sec:Steady_state}

To explore the behavior of the distribution above numerical simulations were performed for the translation of mRNAs incorporating ribosome recycling step using circular boundary condition on a pure initiation state, $V[0]=\{1,0,0…,0\}$ at $t_0=0$. The distribution of proteins was obtained as the difference in fluxes between the termination and re-initiation sites for successively increasing time (Figure~\ref{fig:SteadyState} (a)-(c)), until such time as the steady state distribution of proteins is reached (Figure~\ref{fig:SteadyState} (d)).  The form of protein distribution remains unchanged with any further increase in time, which confirms the asymptotic nature of this distribution. 

\begin{figure}[t]
\begin{centering}
\includegraphics[width=1.0\columnwidth]{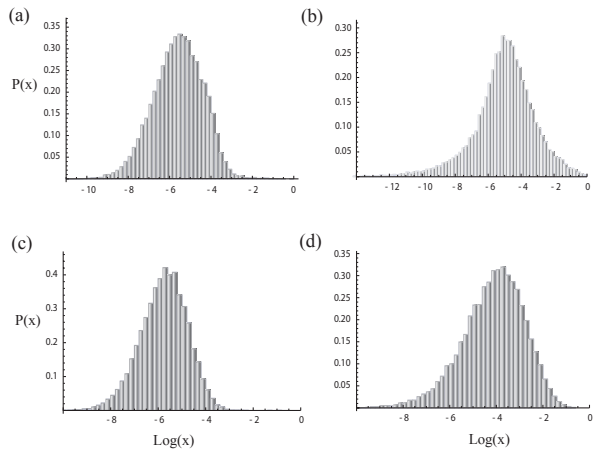} 
\par
\end{centering}
\setlength{\abovecaptionskip}{-60pt plus 3pt minus 2pt}
\caption{\label{fig:EpsilonIndependence} The steady state distribution of proteins does not depend on the precise form of the underlying rate constant distributions for elongation. The plots show the simulated results for the steady state protein distributions in cases in which the underlying rate constant distributions along the entire chain follow (a) normal distribution (mean 50 and standard deviation 15), (b) exponential distribution (mean 100), (c) gamma distribution (shape parameter 10, scale parameter 5), and (d) log-normal distribution (derived from normal distribution with mean 3.5 and standard deviation 1). The steady state distribution of proteins in all cases remains well described by the log-normal distribution. The parameters of the scaled rate constant distribution are chosen so that the system is in the steady state phase in all plots.}
\end{figure}

The effect of the presence of rare codons in the mRNA was also explored using this model. This circumstance should have an effect that is equivalent to a rate limiting phase in the elongation cycle. A linear lattice of size 30 was chosen. At site 20 the scaled rate constant was set close to zero to represent the presence of a rare codon at that site. The calculation of the protein probability density in this condition confirms a local maximum around site 20, consistent with the expectation that the presence of a rare codon on mRNA pauses the translation leading to accumulation near the site (Figure~\ref{fig:RibosomePausing}). The results are similar when the sites with rare codons are chosen near any arbitrary set of consecutive sites on the lattice. Such pausing and stacking effect has been reported in many different experiments, e.g. by Wolin and Walters \citep{Wolin1988}.

\begin{figure}[t]
\begin{centering}
\includegraphics[width=1.0\columnwidth]{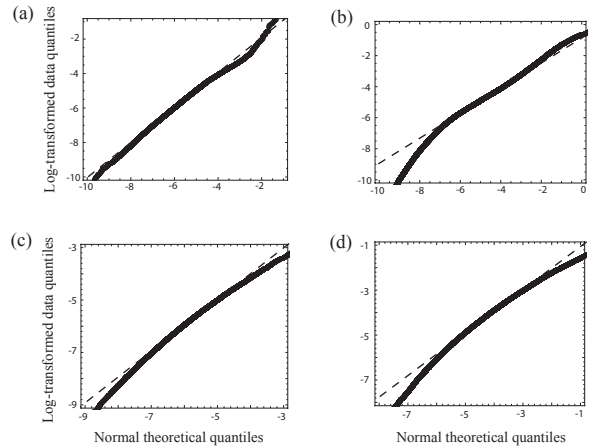} 
\par
\end{centering}
\setlength{\abovecaptionskip}{-60pt plus 3pt minus 2pt}
\caption{\label{fig:qqPlot} A quantile-quantile (Q-Q) plot shows that the steady state protein distribution is well described by a log-normal. The log-transformed steady state distributions are plotted against a normal distribution. The underlying rate constant distributions for the steady state distributions in the Q-Q plots possess the same parameter values as in Figure~\ref{fig:EpsilonIndependence}, given by (a) normal distribution, (b) exponential distribution, (c) gamma distribution, and (d) log-normal distribution.}
\end{figure}

The numerically obtained shape of the steady state distributions in Figure~\ref{fig:SteadyState} resembles a log-normal distribution, which can be explained from the stochastic model outlined in the previous section and described in more detail in the Appendix A. The matrix action $V[t]_i=Q[t]_{i,j} V[0]_j$ on a pure initiation state at $t_0=0$, (i.e. $V[0]=\{1,0,0…,0\}$), results in elements of $V[t]_i$ determined entirely by $Q[t]_{i,1}$, which is given by Eq.~(\ref{eqm3}) for the circular mRNAs and Eq.~(\ref{eqm4}) for the short lived mRNAs. In both these cases $Q[t]_{i,1}$ is a sum of products of random variables. If the second moment of the logarithm of such variables is finite, the product of variables approaches a log-normal distribution as the number of variables grows large, and in such limit $Q[t]_{i,j}$ represents a sum of log-normal distributions. An analytical form for the sum of log-normal distributions cannot be determined as the characteristic function does not have a closed form. But numerical and analytical studies, particularly in the context of wireless communication and related fields, where log-normal sums appear frequently, have shown that the sum of log-normal distributions has similar character to a log-normal distribution \citep{Beaulieu2004} (see Appendix C for detailed discussion). Consistent with this observation we find that the steady state distribution of proteins follows a log-normal, as shown in Figure~\ref{fig:SteadyState}. The invariance of the log-normal distribution under sum implies that the stochastically produced log-normally distributed proteins in single cells when summed over a cell population should also give rise to approximately log-normal distributions in the steady state. 

Broadly, our simulation results are consistent with the emergence of log-normality whenever the range of rate distributions for individual elongation steps remains as large between cells as between individual transcripts within the same cell. It is difficult to plausibly formulate circumstances under which this would not be true.  

\begin{figure}[t]
\begin{centering}
\includegraphics[width=1.0\columnwidth]{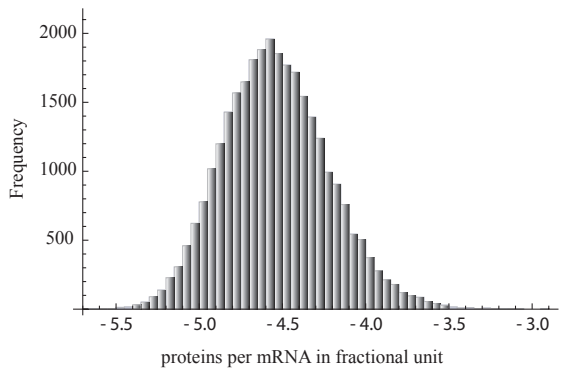} 
\par
\end{centering}
\setlength{\abovecaptionskip}{-80pt plus 3pt minus 2pt}
\caption{\label{fig:ProteinsPerTranscript_Gamma} Number of proteins per mRNA follows a log-normal distribution. The plot shows the histogram of the protein abundance (in fractional units) per template. For the simulation of this plot, the underlying rate constant distribution is taken to follow a gamma distribution with shape parameter 10 and scale parameter 5. Similar results are obtained in cases in which the underlying rate constant distributions are different.}
\end{figure}

An essential tenet of the law of large numbers is the independence of the limiting distribution of sums of variables upon the distributions of the individual variables. Currently, accurate experimental data are not available for rate constant distributions in vivo. Based on chemical reaction kinetics a case can be made that the rate constant distribution may often have an exponential form (See Appendix B). However, as living cells exist in conditions that are far from equilibrium and subject to regulatory influences the possibility that rate constant distributions assume some other form cannot be ruled out. In numerical simulations we have found that the steady state protein distribution form is largely unaffected by the changes in distribution of rate constants, as shown in Figure~\ref{fig:EpsilonIndependence}.

In order to confirm that the asymptotic form of steady state distribution follows a log-normal, we calculated the quantile-quantile plot (Q-Q plot) of the log-transformed distribution against a normal distribution. Figure~\ref{fig:qqPlot} shows that the log-transformed distributions exhibit normality over a wide range, with small deviations seen near the tails. 

Elongation rates may fluctuate over time, but are likely to be slowly varying compared to the time for completion of a polypeptide chain except in unusual circumstances. In Appendix D, we describe two alternative frameworks for numerical calculation for this case and present evidence that the resulting protein distributions are well described by log-normal distribution (Figure~\ref{fig:TimeDependent1} and Figure~\ref{fig:TimeDependent2}). We also show that for low copy number mRNA templates the corresponding protein distribution becomes a Gamma distribution when appropriate limits are applied to the model (Appendix A).

\section{Distribution of the number of polypeptides per transcript}
\label{sec:Polypeptides_transcript}

The number of polypeptides per transcript can be calculated with some assumptions about the decay kinetics of the transcripts. The rate of production of a completed polypeptide of length $d$ is given by $\epsilon_d V_d (t)$ and the integral with respect to time over the lifetime of the transcript gives the number of polypeptides. For first order decay of transcripts with rate constant $c$, the distribution representing the location of the leading ribosome will be the Laplace transform in time of the transition operators with respect to conjugate variable $c$. Taking the example of Eq.~(\ref{eqm3}), the $i,j$ entry of the matrix for $i \geq j$ gives 

\begin{equation}
\label{eqm6}
\frac{\prod_{k=1}^{j-1} (c + \epsilon_k)\prod_{k=i+1}^{d} (c + \epsilon_k) \prod_{m=j}^{i-1} \epsilon_m}{\prod_{k=1}^{d}(c + \epsilon_k)-\lambda \prod_{m=1}^d \epsilon_m}
\end{equation}

Where n is the number of codons as in Eq.~(\ref{eqm3}). The structure of (6) shows the characteristic multiplicative interactions that contribute in an important way to the overall stochastic variation. The consequences of this multiplicative effect can be seen in Figure~\ref{fig:ProteinsPerTranscript_Gamma} which shows that the number of polypeptides per transcript calculated via simulation of Eq.~(\ref{eqm6})  produce a distribution with approximate log-normality. 

\begin{figure}[t]
\begin{centering}
\includegraphics[width=1.2\columnwidth]{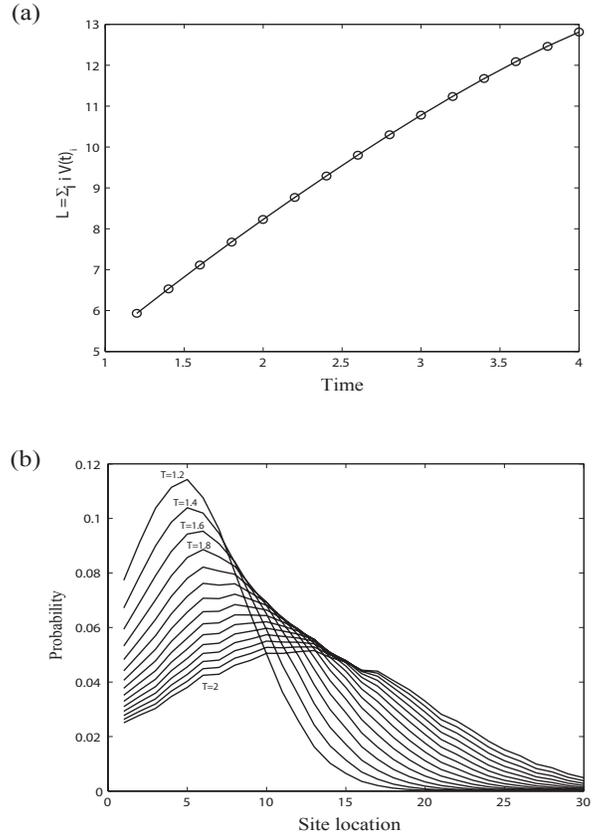} 
\par
\end{centering}
\setlength{\abovecaptionskip}{-30pt plus 3pt minus 2pt}
\caption{\label{fig:ProbabilityInTime} (a) The average length of polypeptide chain d increases linearly with time.  (b) With increasing time the overall occupation probability of polypeptides decreases near the initiation site and increases near the termination site. L denotes length of the lattice chain and m is the mean value of the underlying rate constant distribution, which is given by an exponential distribution for this figure.}
\end{figure}

\begin{figure}[t]
\begin{centering}
\includegraphics[width=1.2\columnwidth]{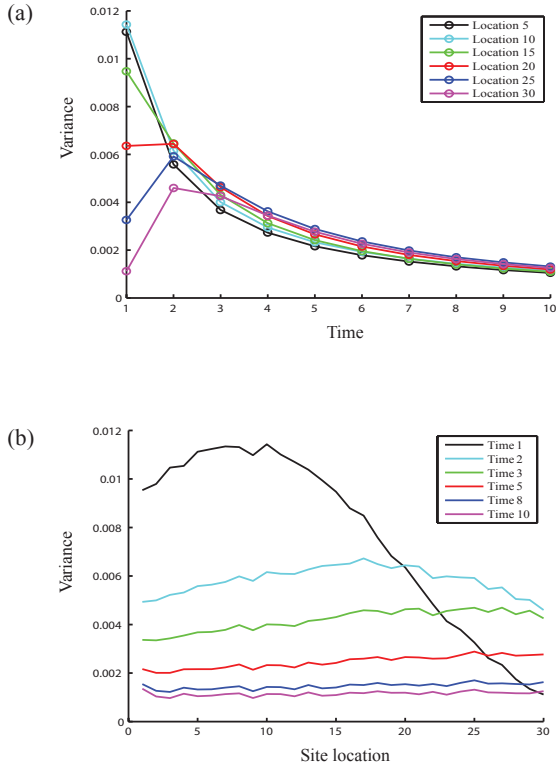} 
\par
\end{centering}
\setlength{\abovecaptionskip}{-30pt plus 3pt minus 2pt}
\caption{\label{fig:Variance} The change in estimated variance of  the occupation probability at different sites along the mRNA. (a) The variation of the estimated variance at different residues along the mRNA chain with time. (b) The changes in estimated variance at different times along the mRNA chain at different residue locations. The estimated variance is obtained by taking the log-transform of a log-normal distribution and calculating the variance of the resulting normal distribution.}
\end{figure}

\section{The transient phase of translation dynamics }
\label{sec:Transient_dynamics}

In this section we closely examine the effect of stochastic dynamics on a collection of mRNAs arising out of a transcriptional burst in a cell population. One useful quantity that quantifies the transient phase of the dynamics is the average extent of polypeptide chain formation in a cell population. We looked at how this quantity varies in time starting from the beginning of the translation process. Let $L$ be such a quantity described by the expectation value

\begin{equation}
\label{eqm7}
L=\sum\limits_{i=1}^{d} i V(t)_i
\end{equation}

Where $i$ denotes the discrete site location and $V(t)_i$ gives the probability that elongation has proceeded up to site i at time t assuming the polypeptide has been initiated at time 0. In Figure~\ref{fig:ProbabilityInTime} (a), we plot $L$ for a lattice chain of length $d=30$ with exponentially distributed rate constants of scaled mean 10. The expected length of the polypeptide chain formed in the cell population, $L$, varies linearly with time, except for a slight deviation from linearity at longer times. To mimic the actual process, in the simulation we allowed the fully formed proteins to decay with some probability. The mass of protein that decays after termination is proportional to the total amount of fully formed protein. The protein decay feature causes a deviation from linearity for $L$ at longer times. A look at the sum of site-wise probabilities across cell population with increasing times (Figure~\ref{fig:ProbabilityInTime} (b)) reveals the effect of elongation proceeding towards the termination site. As expected, with time the occupation probability decreases near the initiation site and increases near the termination site as more and more peptides are released as fully formed proteins.

The variance of the occupation probabilities at different sites across a cell population along the mRNA chain is expected to change with time, as should the occupation probability variance at a particular site. In order to capture the nature of this variation we simulated the process with exponentially distributed spatially varying rate constants over a linear chain of length 30 (Figure~\ref{fig:Variance}
).  With increasing time the numerically estimated variance at different locations tends to converge (Figure~\ref{fig:Variance} (a)) and decrease (Figure~\ref{fig:Variance} (b)).

\section{Protein distribution tends to log-normality except for very low copy number proteins}
\label{sec:Conclusion}

Many physical and chemical laws are multiplicative rather than additive and are expected to lead to log normality in natural systems \citep{Limpert2001}. The log-normal character of protein abundance is often demonstrated by quantitative flow cytometry, in which fluorophore labeled antibodies are reacted with protein targets and the intensity of fluorescence determined on a cell-by-cell basis. The results from flow cytometry, which typically measures cell surface proteins, have been corroborated by results from quantitative mass spectrometry of protein fragments, which also show a log-normal probability density \citep{Boehm2007}.    Various other methods for protein quantitation have led to similar conclusions \citep{Boehm2007, Jung2005, Jung2006, Kaneko2008}.  Nearly all microarray analyses use statistical tests based on the logarithm of raw transcript abundance data \citep{Baldi2002}, consistent with single cell gene expression measurements indicating that mRNA abundance distributions are log-normally distributed in cell populations \citep{Bengtsson2005, Shiku2010}. In this study we have provided a simplified dynamical framework that provides a direct physical explanation of generically observed log-normal distributions. In the limit in which the variation in elongation rates becomes small, the Poisson/Gamma distribution is recovered.

 The deviations in the tails of the distribution in our model indicate that protein abundance distributions in vivo may possess a larger dynamic range than log-normal distributions. These deviations may pose special challenges for cellular regulation if not accompanied by rapid regressions to the mean. At the same time, in a multicellular organism, cells that are outliers with respect to expression may serve a protective or sentinel function, providing a population responses that may have a greater dynamic range attributable to the response characteristics of outliers. 

Considerable work has been done on various aspects of the origin of stochastic gene expression in cellular processes \citep{Elowitz2002, Rodnina2001, Lestas2010, Hilfinger2011}. This work draws attention to the significant and likely dominant role played by translational noise in stochastic gene expression in living cells. Progress in single cell measurements of translational parameters will undoubtedly enhance our understanding of stochastic gene expression.

\appendix

\section{Contour integral form for matrix power series}

A consistent and general representation of the terms of the power series expansion of  $T^k = (U-1)^k$ can be obtained in the form of a contour integral.  In general the matrix $T(d)^k$ with elements $i,j$ can be represented by an integral
\begin{equation}
\label{eqn1}
T(d)^{k}_{i,j}=\frac{1}{2\pi i} \oint g_{i,j}(s,d)s^k ds
\end{equation}
where $g_{i,j} (s,d)$ is a rational function (i.e. a function $f(s)=\frac{p(s)}{q(s)}$, $p, q$ both polynomials of finite order). For integrals of such functions, the sum of the residue at infinity plus the sum of the residues at the zeroes of $q$ equals zero. It is convenient for the purpose of proof by induction to transform the integrand, recalling that the residue at infinity of $f(s)$ is defined as the negative of the residue at 0 of $z^{-2} f(1/z)$, where $z=s^{-1}$. 

In eukaryotic cells evidence of mRNA template circularization has been observed, both biochemically in the form of protein complexes that bind to both poly(A) and the mRNA cap structure \citep{Wells1998}, and ultrastructurally, in the form of polysomes linked in circular configuration. To model the motion of ribosomes on such a template, we identify the generator for the translation operator with the structure 

\begin{equation}
\label{eqn2}
T =
 \begin{pmatrix}
  -\epsilon_1 & 0 & \cdots & 0 & \lambda\epsilon_d \\
  \epsilon_1 & -\epsilon_2 & \cdots & 0 & 0 \\
  \vdots  & \vdots  & \ddots & \vdots & \vdots \\
  0 & 0 & \cdots & -\epsilon_{d-1} & 0 \\
  0 & 0 & \cdots & \epsilon_{d-1} & -\epsilon_d
 \end{pmatrix}
\end{equation} 

from which we construct the desired solution as the sum $\sum_{n=0}^{\infty}(tT)^{n}/n!$. The $i,j$ entry of the $n^{th}$ power of this matrix takes the form, for $i<j$,
\begin{equation}
\label{eqn3}
\frac{1}{2\pi i}\oint\frac{\lambda \prod_{k=i+1}^{j-1}(1+z\epsilon_k) \prod_{m=j}^{d} \epsilon_m \prod_{m=1}^{i-1}\epsilon_m}{z^{j-i+n+1-d}(\prod_{k=1}^{d}(1+z\epsilon_k)-\lambda z^d \prod_{m=1}^d \epsilon_m)}dz
\end{equation}
and
\begin{equation}
\label{eqn4}
\frac{1}{2\pi i}\oint\frac{\prod_{k=1}^{j-1}(1+z\epsilon_k) \prod_{k=i+1}^{d} (1+z\epsilon_k) \prod_{m=j}^{i-1}\epsilon_m}{z^{j-i+n+1}(\prod_{k=1}^{d}(1+z\epsilon_k)-\lambda z^d \prod_{m=1}^d \epsilon_m)}dz
\end{equation}
for $i\geq j$, where the contour of integration encloses the origin but none of the roots of the denominator polynomial  $\prod_{k=1}^{d} (1+z \epsilon_{k})-\lambda z^d \prod_{m=1}^{d} \epsilon_{m}$ . To establish this by induction we first observe that the variable $n$ appears only in the denominator in the exponent of $z$. To calculate the residues at zero we need consider five cases for $n=1$: (i) $i=1,j=d$; (ii) $i< j$ (excepting case (i)); (iii) $i=j$; (iv) $i=j+1$; and (v) $i>j+1$. For case (i), $z^{j-i+2-d}=z$ and the limit of Eq.~(\ref{eqn3}) as $z \rightarrow 0$ is
\begin{equation}
\label{eqn5}
\frac{1}{2\pi i}\oint\frac{\lambda\epsilon_d}{z(1+z\epsilon_1)(1+z\epsilon_d)}dz=\lambda \epsilon_{d}
\end{equation}
For case (ii), the limit of Eq.~(\ref{eqn3}) as $z \rightarrow 0$ is dominated by $j-i+2-d \leq 0$ and the function is analytic within the contour:
\begin{equation}
\label{eqn6}
\frac{1}{2\pi i}\oint\frac{\lambda \prod_{m=j}^{d}\epsilon_m \prod_{m=1}^{i-1}\epsilon_m}{z^{j-i+2-d}}dz=0
\end{equation}
For case (iii) the limit of Eq.~(\ref{eqn4}) as $z \rightarrow 0$ is
\begin{equation}
\label{eqn7}
\frac{1}{2\pi i}\oint\frac{1}{z^{2}(1+z\epsilon_i)}dz \rightarrow \frac{1}{2\pi i}\oint\bigg(\frac{1}{z^{2}}-\frac{\epsilon_i}{z}\bigg)dz=-\epsilon_{i}
\end{equation}
For case (iv) the limit of Eq.~(\ref{eqn4}) as $z \rightarrow 0$ is
\begin{equation}
\label{eqn8}
\frac{1}{2\pi i}\oint\frac{\epsilon_j}{z(1+z\epsilon_j)(1+z\epsilon_{j+1})}dz=\epsilon_{j}
\end{equation}
And for case (v) the limit of Eq.~(\ref{eqn4}) as $z \rightarrow 0$ is dominated by $j-i+2 \leq 0$ and the function is analytic within the contour:
\begin{equation}
\label{eqn9}
\frac{1}{2\pi i}\oint\frac{\prod_{k=j}^{i-1}\epsilon_k}{z^{j-i+2}}dz=0
\end{equation}
To complete a proof by induction we need to formally establish, using Eq.~(\ref{eqn2})~--~Eq.~(\ref{eqn4}) that $T T^n=T^{n+1}$. The actual process is slightly different; we establish that
\begin{equation}
\label{eqn10}
T T^n-T^{n+1}=-\frac{1}{2\pi i}\oint \mathbb{1}_{d}{z^{-n-2}}dz=\mathbb{0}_d
\end{equation} 
Where $\mathbb{1}_{d}$ and $\mathbb{0}_{d}$ are the $d$ dimensional unit and null matrices, respectively. There are 5 cases to be calculated: (i) $i < j ~(i = 1)$; (ii) $i < j ~(i >1)$; (iii) $i = j ~(i = 1)$; (iv) $i = j ~(i >1)$; and (v) $i > j$.  For case (i) we establish
\begin{align}
\label{eqn11}
&\lambda\epsilon_{d} \frac{\prod_{k=1}^{j-1}(1+z\epsilon_{k})\prod_{m=j}^{d-1}\epsilon_{m}}{z^{j-d+n+1}\big(\prod_{k=1}^{d}(1+z\epsilon_{k})-\lambda z^{d} \prod_{m=1}^{d}\epsilon_{m}\big)} \nonumber \\
&-(z^{-1}+\epsilon_{1})
\frac{\prod_{k=2}^{j-1}(1+z\epsilon_{k})\lambda\prod_{m=j}^{d}\epsilon_{m}}{z^{j-d+n}\big(\prod_{k=1}^{d}(1+z\epsilon_{k})-\lambda z^{d} \prod_{m=1}^{d}\epsilon_{m}\big)}=0
\end{align}
For case (ii), 
\begin{align}
\label{eqn12}
&\epsilon_{i-1} \frac{\lambda\prod_{k=i}^{j-1}(1+z\epsilon_{k})\prod_{m=j}^{d}\epsilon_{m}\prod_{m=1}^{i-2}\epsilon_{m}}{z^{j-i+n+2-d}\big(\prod_{k=1}^{d}(1+z\epsilon_{k})-\lambda z^{d} \prod_{m=1}^{d}\epsilon_{m}\big)} \nonumber \\
&-(z^{-1}+\epsilon_{i})
\frac{\prod_{k=i+1}^{j-1}(1+z\epsilon_{k})\lambda\prod_{m=j}^{d}\epsilon_{m}\prod_{m=1}^{i-1}\epsilon_{m}}{z^{j-i+n+1-d}\big(\prod_{k=1}^{d}(1+z\epsilon_{k})-\lambda z^{d} \prod_{m=1}^{d}\epsilon_{m}\big)}=0
\end{align}
For case (iii)
\begin{align}
\label{eqn13}
&\lambda\epsilon_{d} \frac{\prod_{m=1}^{d-1}\epsilon_{m}}{z^{n-d+2}\big(\prod_{k=1}^{d}(1+z\epsilon_{k})-\lambda z^{d} \prod_{m=1}^{d}\epsilon_{m}\big)} \nonumber \\
&-(z^{-1}+\epsilon_{i})
\frac{\prod_{k=2}^{d}(1+z\epsilon_{k})}{z^{n+1}\big(\prod_{k=1}^{d}(1+z\epsilon_{k})-\lambda z^{d} \prod_{m=1}^{d}\epsilon_{m}\big)}=-z^{-n-2}
\end{align}
For case (iv)
\begin{align}
\label{eqn14}
&\epsilon_{i-1} \frac{\lambda\prod_{k=i}^{j-1}(1+z\epsilon_{k})\prod_{m=j}^{d}\epsilon_{m}\prod_{m=1}^{i-2}\epsilon_{m}}{z^{j-i+n+2-d}\big(\prod_{k=1}^{d}(1+z\epsilon_{k})-\lambda z^{d} \prod_{m=1}^{d}\epsilon_{m}\big)} \nonumber \\
&-(z^{-1}+\epsilon_{i})
\frac{\prod_{k=1}^{j-1}(1+z\epsilon_{k})\prod_{k=i+1}^{d}(1+z\epsilon_{k})\prod_{k=j}^{i-1}\epsilon_{m}}{z^{j-i+n+1}\big(\prod_{k=1}^{d}(1+z\epsilon_{k})-\lambda z^{d} \prod_{m=1}^{d}\epsilon_{m}\big)}=-z^{-n-2}
\end{align}
And for case (v)

\begin{align}
\label{eqn15}
&\epsilon_{i-1} \frac{\prod_{k=1}^{j-1}(1+z\epsilon_{k})\prod_{k=i}^{d}(1+z\epsilon_{k})\prod_{m=j}^{i-2}\epsilon_{m}}{z^{j-i+n+2}\big(\prod_{k=1}^{d}(1+z\epsilon_{k})-\lambda z^{d} \prod_{m=1}^{d}\epsilon_{m}\big)} \nonumber \\
&-(z^{-1}+\epsilon_{i})
\frac{\prod_{k=1}^{j-1}(1+z\epsilon_{k})\prod_{k=i+1}^{d}(1+z\epsilon_{k})\prod_{m=j}^{i-1}\epsilon_{m}}{z^{j-i+n+1}\big(\prod_{k=1}^{d}(1+z\epsilon_{k})-\lambda z^{d} \prod_{m=1}^{d}\epsilon_{m}\big)}=0
\end{align}
A more convenient characterization of Eq.~(\ref{eqn3}) and Eq.~(\ref{eqn4}) for evaluation of the steady state formulates the matrix in terms of $s=z^{-1}$ and gives, for the $i,j$ entry of the $n^{th}$ power of the matrix Eq.~(\ref{eqn1}), for $i<j$
\begin{equation}
\label{eqn16}
\frac{1}{2\pi i}\oint\frac{s^{n}\lambda\prod_{k=i+1}^{j-1}(s+\epsilon_{k})\prod_{m=j}^{d}\epsilon_{m}\prod_{m=1}^{i-1}\epsilon_{m}}{\prod_{k=1}^{d}(s+\epsilon_{k})-\lambda \prod_{m=1}^{d}\epsilon_{m}}ds
\end{equation}
and
\begin{equation}
\label{eqn17}
\frac{1}{2\pi i}\oint\frac{s^{n}\prod_{k=1}^{j-1}(s+\epsilon_{k})\prod_{k=i+1}^{d}(s+\epsilon_{k})\prod_{m=j}^{i-1}\epsilon_{m}}{\prod_{k=1}^{d}(s+\epsilon_{k})-\lambda \prod_{m=1}^{d}\epsilon_{m}}ds
\end{equation}
otherwise, where the contour encloses all of the roots of the denominator polynomial $\prod_{k=1}^{d}(s+\epsilon_k)-\lambda \prod_{m=1}^{d}\epsilon_{m}$.  For $\lambda=1$, $T$ is a stochastic (conservative) matrix, and therefore should have a steady state given by the residues of Eq.~(\ref{eqn16}) and Eq.~(\ref{eqn17}) at $s=0$. Inspection of Eq.~(\ref{eqn16}) and Eq.~(\ref{eqn17}) shows that for $\lambda=1, s=0$ is indeed a root of the denominator polynomial, and hence the steady state solution is given by the residue at 0 of the sum of Eq.~(\ref{eqn16}) and Eq.~(\ref{eqn17}) over $n$ with $s^n$ replaced by $(ts)^{n}/n!$ , which is 
\begin{equation}
\label{eqn18}
\frac{\prod_{m=1}^{d}\epsilon_{m}}{\epsilon_{i}\mathlarger{\mathlarger{\sum_{k=1}^{d}}}\mathlarger{\frac{\prod_{m=1}^{d}\epsilon_{m}}{\epsilon_{k}}}}=\frac{1}{\epsilon_{i}\mathlarger{\sum_{k=1}^{d}\frac{1}{\epsilon_{k}}}}
\end{equation}
in both cases. The $i,j$ entry of the matrix representing the steady state distribution has no dependence on $j$, consistent with intuition.

In the case $\lambda=0$, the $i,j$ entry of the $n^{th}$ power of matrix (\ref{eqn2}) is zero for $i<j$, and
\begin{equation}
\label{eqn19}
\frac{1}{2\pi i}\mathlarger{\oint}\frac{s^{n}\prod_{m=j}^{i-1}\epsilon_m}{\prod_{k=j}^{i}(s+\epsilon_{k})}
\end{equation}
otherwise, where the contour taken in the conventional (positive) sense encloses all of the roots of the denominator polynomial (i.e. encircles all of the real axis values of $-\epsilon_{k},j \leq k \leq i$. The matrix $e^{t T}$ in this case is given by Eq.~(\ref{eqn19}) with $s^n$ replaced by the sum $\sum_{n=0}^{\infty}(ts)^n/n!$, the integral of which converges, despite its resemblance to a function with an essential singularity at infinity. We note also for completeness that the evaluation of the contour integral for $n = 0$ yields the unit matrix as needed. Evaluation of the integral leads to $(e^{t T})_{i,j}=0$ for $i<j$, and
\begin{equation}
\label{eqn20}
(e^{tT})_{i,j}=(-1)^{i+j}\mathlarger{\mathlarger{\sum_{k=j}^{i}}}\frac{e^{-t\epsilon_{k}}\prod_{m=j}^{i-1}\epsilon_{m}}{\prod_{p=1}^{i-k}(\epsilon_{k}-\epsilon_{k+p})\prod_{q=j}^{k-1}(\epsilon_{k}-\epsilon_{q})}
\end{equation}
otherwise.

In the event that all of the $\epsilon_k$ are equal, the evolution operator represented by Eq.~(\ref{eqn20}) takes the particularly simple form 
\begin{equation}
\label{eqn21}
\frac{(t\epsilon)^{i-j}}{(i-j)!}e^{-t\epsilon}
\end{equation}
for $i \geq j$ (and 0 otherwise) and hence Eq.~(\ref{eqn20}) can be considered a natural generalization of a Poisson process to a domain in which the underlying stochastic process is not homogeneous. The gamma distribution, an extension of the Poisson to nonintegral event frequencies, takes the related form 
\begin{equation}
\label{eqn22}
\frac{(t\epsilon)^{k}}{\Gamma(k+1)}e^{-t\epsilon}
\end{equation}
which bears comparison to Eq.~(\ref{eqn20}) because Eq.~(\ref{eqn22}) has been proposed to appropriately capture the statistics of low multiplicity translations emitted by a single mRNA template. 

When the operator $T$ acts on an initial state vector $v(t_0)=(1,0,\ldots,0)$ of length $d$ at time $t_0=0$, $e^{t T} v(0)$ gives the probability density of the location of a ribosome on the mRNA at time $t$.  If the source of the translation is an mRNA with a probability of existence at time $t$ of $c e^{-c t}$, the relative effect of additional initiations will be given by $\int_{0}^{\infty}c e^{-c t} e^{t T} v(0) dt$, and the $i,j$ entry of  $\int_{0}^{\infty}c e^{-c t} e^{t T} dt$ for $i<j$ is 
\begin{equation}
\label{eqn23}
\frac{c\lambda\prod_{k=i+1}^{j-1}(c+\epsilon_{k})\prod_{m=j}^{d}\epsilon_{m}\prod_{m=1}^{i-1}\epsilon_{m}}{\prod_{k=1}^{d}(c+\epsilon_{k})-\lambda \prod_{m=1}^{d}\epsilon_{m}}
\end{equation}
and
\begin{equation}
\label{eqn24}
\frac{c\prod_{k=1}^{j-1}(c+\epsilon_{k})\prod_{k=i+1}^{d}(c+\epsilon_{k})\prod_{m=j}^{i-1}\epsilon_{m}}{\prod_{k=1}^{d}(c+\epsilon_{k})-\lambda \prod_{m=1}^{d}\epsilon_{m}}
\end{equation}
otherwise.

The rate of production of full length protein is given by $v(t)_d \epsilon_{d}=e^{t T} v(0)_d \epsilon_{d}=(e^{t T})_{d,1} \epsilon_{d}$ and the integral with respect to time weighted by the lifetime of the encoding RNA gives
\begin{equation}
\label{eqn25}
c e_{d}\int_{0}^{\infty}e^{-c t}(e^{t T})_{d,1}dt=\frac{c \prod_{m=1}^{d}\epsilon_{m}}{\prod_{k=1}^{d}(c+\epsilon_{k})-\lambda \prod_{m=1}^{d}\epsilon_{m}}
\end{equation}
for the average number of polypeptides produced per mRNA template, assuming that the characteristic lifetime of the mRNA, $c^{-1}$ , is long compared to the translation time. In the event this is not true, we can estimate the number of polypeptides per template in the elongation-limited domain by dividing the mean length that the lead ribosome has translated down the mRNA, divided by the average number of residues between successive ribosomes, $R$. This has the form
\begin{equation}
\label{eqn26}
\mathlarger{\sum_{i=1}^{d}}\frac{i}{R}\int_{0}^{\infty}c e^{-c t}v(t)_{i}dt=\mathlarger{\sum_{i=1}^{d}}\frac{i}{R}\int_{0}^{\infty}c e^{-c t}(e^{t T})_{i,1}dt
\end{equation}
which, using Eq.~(\ref{eqn23}) and setting $\lambda=0$ (since the probability of reinitiation can be neglected), gives
\begin{equation}
\label{eqn27}
\mathlarger{\sum_{i=1}^{d}}\frac{i}{R}\frac{c \prod_{m=1}^{i-1}\epsilon_{m}}{\prod_{k=1}^{i}(c+\epsilon_{k})}
\end{equation}
for the mean number of ribosomes per template over the life of the template. In the limit that all $\epsilon_k$ are equal, Eq.~(\ref{eqn25}) gives
\begin{align}
\label{eqn28}
\mathlarger{\mathlarger{\sum_{i=1}^{d}}}\frac{i}{R}\frac{c}{\epsilon~\mathlarger{\bigg(1+\frac{c}{\epsilon}\bigg)^{i}}}
&=\frac{(c+\epsilon)~\bigg(1-\mathlarger{\bigg(\frac{c+\epsilon}{\epsilon}~\bigg)^{-d}}\bigg)-cd~\mathlarger{\bigg(\frac{c+\epsilon}{\epsilon}\bigg)^{-d}}}{Rc} \nonumber \\
&\approx\frac{(c+\epsilon)~\bigg(1-e^\mathlarger{{-\frac{cd}{\epsilon}}}\bigg)-cd~ e^\mathlarger{{-\frac{cd}{\epsilon}}}}{Rc}
\end{align}
the latter approximation holding for $c \ll \epsilon$.

\section{Rate constant distributions}
The individual $\epsilon_i$ as modeled here are lumped rate constants for chemical reactions of considerable complexity. However even complex trajectories in reaction coordinates can often be modeled by taking the reaction to proceed through a limiting intermediate corresponding to the lowest energy barrier of the transition state, which by convention ascribes to the $\epsilon_i$ a structure
\begin{equation}
\label{eqnB1}
\epsilon_{i}=A_{i}~ \mathlarger{e^{-\frac{\Delta G_{i}}{RT}}}
\end{equation}
where $\Delta~G_i$ is the Gibbs free energy for the $i$th transition state, $R$ is the universal gas constant, $T$ is the temperature in absolute scale, and $A_i$ is a constant. The transition state free energy is defined in terms of the corresponding enthalpy ($H$) and entropy ($S$) in the usual way as
\begin{equation}
\label{eqnB2}
\Delta G_i=\Delta H_i-T \Delta S_i		
\end{equation}
We assume that the correctly charged tRNA reaches the ribosome A site through a diffusion process which in turn determines the rate at which the elongation phase proceeds. Since ordinary diffusion is dominated by entropic contributions, we assume the enthalpic contribution can be neglected, and the rate constant is effectively given by 
\begin{equation}
\label{eqnB3}
\epsilon_{i}=A_{i}~ \mathlarger{e^{-\frac{\Delta S_{i}}{R}}}
\end{equation}
If we let $\omega_i$ represents the number of microstates corresponding to the macrostate of the system, Eq.~(\ref{eqnB3}) can be further modified as follows
\begin{equation}
\label{eqnB4}
\epsilon_{i}=A_{i}~\mbox{exp}~\bigg(\frac{k_{B}~\Delta ~ln ~\omega_{i}}{Nk_{B}}\bigg)
=A_{i}~\mbox{exp}~\bigg(\frac{\Delta ~ln~ \omega_{i}}{N}\bigg)
\end{equation}
where $k_{B}$ is Boltzmann’s constant and $N$ Avogadro’s number. The change in entropy depends on the difference in configurations between the microstates of transitioning states and is difficult to ascertain precisely when the states between which the transition occurs are both far from equilibrium. However our numerical calculation suggests that universality in the steady state distribution of protein probability density, the counterpart of the law of large numbers in the setting of multiplicative variables, has a rapid onset, such that even very short polypeptides (less than 30 residues) show universal behavior. In numerical simulations, in addition to the exponential distribution we have used the Gamma, normal, log-normal and uniform distribution as possible forms of forms for the rate constant distribution (Figure~\ref{fig:EpsilonIndependence}). 

\section{Sums of log-normally distributed variables generate distributions that behave similarly to log-normal variables}
The steady state distribution of proteins is given by a sum of the products of random variables. The sum of independent and identically distributed random variables with finite variance tends to a normal distribution as the number of variables grows large. Similarly, the distribution of a product of random variables converges to a log-normal distribution as the number of terms of the product increases. But the calculation of the sum of log-normal distributions themselves faces a theoretical roadblock. The distribution of a sum of independent random variables is obtained from the product of the respective characteristic functions, but for the log-normal distribution a closed form for the characteristic function does not exist and the sum of log-normally distributed variates has not been  obtained in a closed form\citep{Fenton1960}.

The behavior of the sums of log-normally distributed variables have been a topic of interest in field of communications for over half a century; specific examples where sums of log-normals appear include co-channel interference in mobile (wireless) communications, in frequency hopped spread spectrum signals and in the general context of propagation through turbulent medium \citep{Beaulieu2004,Beaulieu1995}.  Extensive numerical evidence from these studies have confirmed that the sum of independent log-normally distributed random variables is well approximated by a log-normal distribution \citep{Beaulieu2004,Fenton1960,Beaulieu1995,Schwartz1982}. We observe similar behavior for the numerically calculated distributions for the steday state distribution of proteins as shown in Figure~\ref{fig:SteadyState} and Figure~\ref{fig:EpsilonIndependence} of the main text. 

\section{Rate constants fluctuating in time}
\begin{figure}[t]
\begin{centering}
\includegraphics[width=1.0\columnwidth]{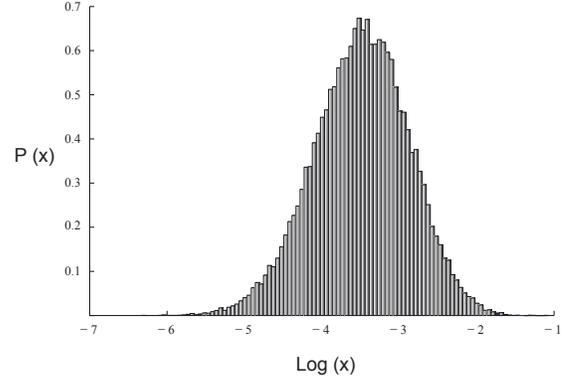} 
\par
\end{centering}
\setlength{\abovecaptionskip}{-70pt plus 3pt minus 2pt}
\caption{\label{fig:TimeDependent1} A simulation result for the steady state distribution of proteins across a cell population when rate constants depend on both location and time (Appendix D). For the simulation, the length of the lattice chain was taken as 30 and the rate constants were drawn from an exponential distribution.}
\end{figure}

So far we have considered the case in which the rate constants are location dependent but constant in time. Here we consider the most general case, in which the transition probability for the state vector passing from state $i$ to state $i+1$,  $\epsilon_{i}^{'}$, also depends on the time of the transition. In this case, we may define a time dependent transition operator $U(t_{i})$  as
\begin{equation}
\label{eqnD1}
U(t_{i}) =
 \begin{pmatrix}
  1-\epsilon_{1}^{'}(t_{i}) & 0 & \cdots & 0 & \lambda\epsilon_d \\
  \epsilon_{1}^{'}(t_{i}) & 1-\epsilon_{2}^{'}(t_{i}) & \cdots & 0 & 0 \\
  \vdots  & \vdots  & \ddots & \vdots & \vdots \\
  0 & 0 & \cdots & 1-\epsilon_{d-1}^{'}(t_{i}) & 0 \\
  0 & 0 & \cdots & \epsilon_{d-1}^{'}(t_{i}) & 1-\epsilon_{d}^{'}(t_{i})
 \end{pmatrix}
\end{equation}
The most general case represented by the above transition matrix where $\epsilon_{i}^{'}$s vary both in space and time, can be obtained by direct numerical simulation. In Figure~\ref{fig:TimeDependent1} we have shown the simulation for the case of exponentially distributed $\epsilon_{i}^{'}$s. The resulting distribution follows a log-normal form consistent with general expectations regarding the universality of this distribution form.

To extract approximate behavior analytically, we consider that the most appropriate approach   will depend on the nature of the time dependence of $\epsilon_{i}^{'}$. The simplest case occurs when the time dependence of $\epsilon_{i}^{'}$ can be represented as a smooth function which is independent of the location $i$. The bi-directional version of this case with a fixed coefficient at all locations has a solution in terms of infinite sums of the modified Bessel function \citep{Kleinrock1975}. The more general version involving location-dependent but smooth functional time dependence (with location dependent coefficients) of $\epsilon_{i}^{'}$ becomes analytically intractable when formal methods such as a power series representation for solving linear coupled ODE system are used.  

Due to the random nature of environmental influences on the translation process, the appropriate form of time dependence of $\epsilon_{i}^{'}$ is stochastic. Typical stochastic functions are not integrable, which makes them weak candidates for calculating time dependence. However, Ito processes, a general class of stochastic functions for which well-developed procedures for stochastic integration exist, can be applied. An Ito process is a diffusion process for which both drift and diffusion rates are functions of time. Therefore in a general mathematically well-founded approach, we can consider the $\epsilon_{i}^{'}$ to be drawn from an Ito process in order to model the influence of random environment on the translation. 

From biological considerations the time dependence of rate constants is most likely to be slowly varying with a narrow distribution range. Under such circumstances, there are simpler and physically more illuminating alternative procedures to Ito process formalism, for obtaining the solution for the time-dependent case. We discuss two such formulations both of which capture the dynamics in the case of slowly varying and narrowly distributed time dependent $\epsilon_{i}^{'}$.

The first procedure involves the power series expansion of the Poisson stochastic operator.  Let the times be ordered so that $t_1<t_2<t_3<\ldots<t_n$. Then the overall transition operator at the time $t_{n}$, $Q(t_{n})$ is given by 
\begin{align}
\label{eqnD2}
Q(t_{n})=&~e^{-\alpha t_{n}}[1+(\alpha t_{n})U(t_{1})+\frac{(\alpha t_{n})^2}{2!}U(t_{2}).U(t_{1})
+\ldots \nonumber \\ &+\frac{(\alpha t_{n})^{n}}{n!}U(t_{n})\ldots U(t_{2}).U(t_{1})]
\end{align}
The individual terms in the power series expansion of Poisson operator are represented as time-ordered products of $U(t_{i})$.  The operators at different time points $U(t_{i})$ can be constructed by drawing $\epsilon_{i}^{'} (t_{i})$’s from a specific distribution, following which the transition operator can be computed directly from the above equation. In the case of the translation process the time dependence of $\epsilon_{i}^{'} (t_{i})$'s are likely to be slowly varying with a narrow distribution, and in such cases the $Q(t_{n})$ is guaranteed to converge for some suitably high value of $t_{n}$, which we have observed numerically.

When the rate constants are time dependent, formula Eq.~(\ref{eqnD2}) for the evolution does not in general hold because the exponentiated matrix operator do not commute. We consider the case in which the constants $\epsilon_{i}$ of Eq.~(\ref{eqnD2}) are replaced by functions $\epsilon_{i}(t)$ piecewise constant over sequential epochs $\Delta t_k$ of constant duration $\Delta t$. Then if all $\epsilon_{i}(\Delta t_{k})=\epsilon_{i}(\Delta t_{k+1})$ we have
\begin{equation}
\label{eqnD3}
Q(\Delta t_{k})Q(\Delta t_{k+1})=\mathlarger{e^{\alpha \Delta t G (\Delta t_{k})}e^{\alpha \Delta t G (\Delta t_{k+1})}}
=\mathlarger{e^{2\alpha \Delta t G (\Delta t_{k})}}
\end{equation}
However, if the rate constants vary in time, i.e  $\epsilon_{i}(\Delta t_{k})\neq \epsilon_{i}(\Delta t_{k+1})$,
\begin{equation}
\label{eqnD4}
Q(\Delta t_{k})Q(\Delta t_{k+1})\neq \mathlarger{e^{\alpha \Delta t G (\Delta t_{k})+\alpha \Delta t G (\Delta t_{k+1})}}
\end{equation}
unless the commutator $[G (\Delta t_{k}),G (\Delta t_{k+1})]$ vanishes.
\begin{align}
\label{eqnD5}
&[G (\Delta t_{k}),G (\Delta t_{k+1})]_{i,j} \nonumber \\
&=\mathlarger{\epsilon_{j}(\Delta t_{k})\epsilon_{i}(\Delta t_{k+1})-\epsilon_{i}(\Delta t_{k})\epsilon_{j}(\Delta t_{k+1})} ~~~~~~~~~~~i=j+1\nonumber \\
&=\mathlarger{\epsilon_{i-1}(\Delta t_{k})\epsilon_{j}(\Delta t_{k+1})-\epsilon_{j}(\Delta t_{k})\epsilon_{i-1}(\Delta t_{k+1})} ~~~~i=j+2 \nonumber \\
&=0 ~~~~~~~~~~~~~~~~~~~~~~~~~~~~~~~~~~~~~~~~~~~~~~~~~~~~~~~~~~~~~~\mbox{otherwise}
\end{align}
If all of the columns except the $(n-1)$'{th} sum to 0, then we can formulate the time dependence by dividing the entire time duration in several intervals such that within each interval the rate constants remain constant. 

\begin{figure}[t]
\begin{centering}
\includegraphics[width=1.0\columnwidth]{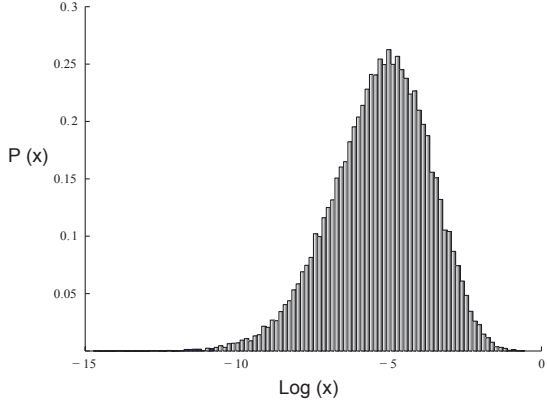} 
\par
\end{centering}
\setlength{\abovecaptionskip}{-70pt plus 3pt minus 2pt}
\caption{\label{fig:TimeDependent2} The simulated steady state distribution of proteins across a cell population. For the simulation, rate constants were allowed to change not only with location but were also slowly varying in time (Appendix D). The length of lattice chain for the simulation was 30 and the rate constants were drawn from an exponential distribution.}
\end{figure}

The second procedure for handling the time dependent transition operator can be formulated by replacing the non-stationary transition operator $U(t_{i})$ with a series of stationary transition operators, simulating the situation in which the rate constants $\epsilon_{i}^{'}$ are slowly varying in time. Let the total time interval over which the dynamical evolution is observed be given by $t$. Suppose that the total time interval $t$ can be divided into $n$ time sub-intervals of arbitrary lengths, such that over each such time sub-intervals the transition operator  $U(t_{i})$ is given by a stationary transition operator, i.e. $t=t_{1}+t_{2}+\ldots+t_{n}$.  The transition matrix $U(t_{i})$, is expressed in terms of the probability vectors $\epsilon_{i,t_{i}}^{'}$, where $\epsilon_{i,t_{i}}^{'}$ is the probability that the state vector moves from state $i$ to state $i+1$ per unit time, anytime during the time interval $t_{i}$. The transition operator can be written as
\begin{equation}
\label{eqnD6}
U(t_{i}) =
 \begin{pmatrix}
  1-\epsilon_{1,t_{i}}^{'} & 0 & \cdots & 0 & 0 \\
  \epsilon_{1,t_{i}}^{'} & 1-\epsilon_{2,t_{i}}^{'} & \cdots & 0 & 0 \\
  \vdots  & \vdots  & \ddots & \vdots & \vdots \\
  0 & 0 & \cdots & 1-\epsilon_{d-1,t_{i}}^{'} & 0 \\
  0 & 0 & \cdots & \epsilon_{d-1,t_{i}}^{'} & 1-\epsilon_{d,t_{i}}^{'}
 \end{pmatrix}
\end{equation}
For the Poisson stochastic process, the transition operator over the entire interval $t$ can now be written as a product of a series of transition operators over each sub-interval $t_{i}$. Therefore we have 
\begin{equation}
\label{eqnD7}
Q(t)=e^{\alpha t_{1}(U(t_{1})-1)}.e^{\alpha t_{2}(U(t_{2})-1)}\ldots e^{\alpha t_{n}(U(t_{n})-1)}
\end{equation}
We can replace $\epsilon_{i,t_{i}}^{'}=\mathlarger{\frac{\epsilon_{1,t_{i}}}{\alpha}}$ and write
\begin{align}
\label{eqnD8}
\alpha(U(t_{i})-1) &\equiv\alpha G(t_{i}) \nonumber \\
&=
 \begin{pmatrix}
  -\epsilon_{1,t_{i}} & 0 & \cdots & 0 & 0 \\
  \epsilon_{1,t_{i}} & -\epsilon_{2,t_{i}} & \cdots & 0 & 0 \\
  \vdots  & \vdots  & \ddots & \vdots & \vdots \\
  0 & 0 & \cdots & -\epsilon_{d-1,t_{i}} & 0 \\
  0 & 0 & \cdots & \epsilon_{d-1,t_{i}} & -\epsilon_{d,t_{i}}
 \end{pmatrix}
\end{align}
So we have
\begin{equation}
\label{eqnD9}
Q(t)=e^{\alpha t_{1}G(t_{1})}.e^{\alpha t_{2}G(t_{2})}\ldots e^{\alpha t_{n}G(t_{n})}
\end{equation}
We have carried out numerical simulation designed to measure the consequences of this form of temporal variation. The resulting distribution has a log-normal shape as shown in Figure~\ref{fig:TimeDependent2}.

If the matrices $e^{\alpha t_{i} G(t_{i})}$ for different time intervals commute with each other then the final form of $Q(t)$ simplifies. To simplify the discussion, let us assume that the time subintervals are all of equal duration given by $\tau$ and $\epsilon_{i,t_{1}}=\omega_i$ and $\epsilon_{i,t_{2}}=\delta_i$. The combined time evolution operator for two successive intervals will be given by  
\begin{equation}
\label{eqnD10}
Q(t)=e^{\alpha \tau G(t_{1})}.e^{\alpha \tau G(t_{2})}
\end{equation}
This can be rewritten as 
\begin{equation}
\label{eqnD11}
Q(t)=e^{\alpha \tau G(t_{1})+\alpha \tau G(t_{2})+\alpha \tau [G(t_{1}),G(t_{2})]}
\end{equation}
Note that if the transition operators $G(t_{1})$ and $G(t_{2})$ commute then the formulae for  $Q(t)$ involve direct addition in the exponential. Consequently, when this commutation condition holds for all successive time intervals then the effective time evolution formulae becomes considerably simpler. The commutator of  $\alpha G(t_{1})$ and $\alpha G(t_{2})$  is given by
\begin{equation}
\label{eqnD12}
[G(t_{1}),G(t_{2})] =
 \begin{pmatrix}
  0 & 0 & \cdots & 0 &\\
  -\omega_{1} \delta_{2}+\omega_{2} \delta_{1} & 0 & \cdots & 0\\
  \omega_{1} \delta_{2}-\omega_{2} \delta_{1}  & -\omega_{2} \delta_{3}+\omega_{3} \delta_{2}  & \ddots & \vdots &\\
  0 & \omega_{2} \delta_{3}-\omega_{3} \delta_{2} & \cdots & \vdots\\
  \vdots & \vdots & \cdots & \vdots
 \end{pmatrix}
\end{equation}
There are two separate conditions on the hopping probabilities under which $\alpha G(t_{1})$ and $\alpha G(t_{2})$ will commute

\begin{itemize}
\item[1)] Since the entries in the commutator matrix are of the order of the square of rate constants, for very small values of rate constant, $\alpha G(t_{1})$ and $\alpha G(t_{2})$ commute.

\item[2)] The entries in the commutator have the general form $-\omega_{i} \delta_{i+1}+\omega_{i+1} \delta_{i}$. If the rate constants change in time in a correlated manner such that $\delta$ becomes a function of $\omega$, then 
\begin{equation}
\label{eqnD13}
-\omega_{i} \delta_{i+1}+\omega_{i+1} \delta_{i}=0
\end{equation}
This in turn implies that $\alpha G(t_{1})$ and $\alpha G(t_{2})$ will commute. 
\end{itemize}

Under either of these conditions, the general formula for the time evolution in the case of space-time dependent rate constants will be given by    
\begin{equation}
\label{eqnD14}
Q(t)=e^{\alpha \tau [G(t_{1})+ G(t_{2})+\ldots + G(t_{n})]}
\end{equation}
The above formula can be easily generalized for the case when the subinterval time durations are not equal to each other, in which case the formula is given by
\begin{equation}
\label{eqnD15}
Q(t)=e^{\alpha [t_{1} G(t_{1})+t_{2} G(t_{2})+\ldots +t_{n} G(t_{n})]}
\end{equation}	
In contrast to the more general formula given by Eq.~(\ref{eqnD9}), in Eq.~(\ref{eqnD15}) the specific time ordering of different sub-intervals becomes unimportant in the overall form of Poisson semi-group transition operator. 

\section{Note on numerical simulation}
For the distribution of elongation rates, the rate constants at various sites are scaled by both $\alpha$ and time, where $\alpha$ has the dimension of inverse time. Once $\alpha$  is fixed at a specific value,  longer time evolution is given by scalar multiplication of the rate constant distribution values by a higher factor. We refer to these resulting rate constant values as scaled rate constants. The numerical simulations presented in the paper were carried out in 32 bit Matlab R2009b and Mathematica 9.

\bibliography{References}

\end{document}